\documentclass{elsart}

\usepackage{epsfig}
\usepackage{amsmath}
\usepackage{amssymb}
\usepackage{latexsym}
\usepackage{exscale}

\renewcommand{\vec}[1]{\mbox{\boldmath{$#1$}}}

\begin{document}

\begin{frontmatter}

\title{Investigation of the performance of an optimised MicroCAT, a GEM 
       and their combination by simulations and 
       current measurements\thanksref{EU}}
\thanks[EU]{Work supported by the European Community
            (contract no. ERBFMGECT980104).}

\author[Siegen]{A. Orthen\corauthref{Autor}},
\author[Siegen]{H. Wagner},
\author[Siegen]{H.J. Besch},
\author[Trieste]{R.H. Menk},
\author[Siegen]{A.H. Walenta},
\author[Siegen]{U. Werthenbach}

\corauth[Autor]{Corresponding author.
Tel.: +49 271-740-3563;
fax:  +49 271-740-3533;
e-mail: orthen@alwa02.physik.uni-siegen.de.}

\address[Siegen]
  {Universit\"at Siegen, Fachbereich Physik,
   Emmy-Noether-Campus,
   Walter-Flex-Str. 3, 57068 Siegen, Germany}
\address[Trieste]
  {ELETTRA, Sincrotrone Trieste, S.S. 14, km 163.5,
   Basovizza, 34012 Trieste, Italy}

\begin{abstract}
A MicroCAT (Micro Compteur \`{a} Trous) structure which is used for
avalanche charge multiplication in gas filled radiation detectors
has been optimised with respect 
to maximum electron transparency and minimum ion feedback.
We report on the charge transfer behaviour and the achievable gas gain
of this device.
A three-dimensional electron and ion transfer simulation
is compared to results derived from
electric current measurements.
Similarly, we present studies of 
the charge transfer behaviour of a GEM (Gas Electron
Multiplier) by current measurements and simulations.
Finally, we investigate the combination of the MicroCAT and the GEM by
measurements with respect to the performance at
different voltage settings, gas mixtures and gas pressures.
\\
\vspace{4mm}
\end{abstract}


\end{frontmatter}

\section{Introduction}

With the rise of micropattern gas gain devices like 
GEM \cite{Sauli}, MICROMEGAS \cite{Giomataris}, 
CAT \cite{Lemonnier}  
or MicroCAT \cite{Amir3} the field of application of gas 
filled detectors has been widened up. 
These structures accept high rates and 
due to their parallel plate geometry they produce short signals and  
good time-resolution.
\\ 
Several attempts have been made in the past 
to enhance the gain and the reliability 
by combining micropattern structures.  
The performance of the combination of MSGCs with GEM structures is
reported in Refs.
\cite{Benhammou,Baruth,Bagaturia}. The combinations of MSGCs and MGCs 
with GEM, MICROMEGAS and Plate Avalanche Chambers
are discussed in Ref. \cite{Fonte} with a special emphasis on 
rate and gain limitations. 
A GEM, used as a preamplification stage, has also been
successfully combined with a Groove Chamber \cite{Reichwein}.
First results of a device combining a 
MICROMEGAS and a GEM have been presented in Ref. \cite{Kane}. 
GEM structures have been combined to 
double GEM \cite{Bachmann}, triple GEM and even to quad GEM 
configurations \cite{Buzulutskov1,Buzulutskov2}, which reach an 
enormous gas gain in the order of $10^5-10^6$.   
\\
Adding a GEM to the MicroCAT structure
a stable operation with respect to sparks 
is obtained at moderate potentials; nevertheless  
quite high gas gains are achievable at higher potentials.  
For this reason the detector can be operated
in applications where higher gas pressures ($\approx3\,\mathrm{bar}$) 
and high-$Z$ gases (like e.g. Xenon) are required, i.e. in X-ray
detection, as will be shown in this paper.
\\
The intention of this investigation is to optimise MicroCAT- and
GEM-based detector systems for X-ray imaging (typical photon energy range: 
$5-25\,\mathrm{keV}$), where the drift
gaps are usually several $\mathrm{cm}$ large and the drift fields are
in the range of $1\,\mathrm{kV/cm}$ or even less.

\section{Study of the optimised MCAT215
\label{MCAT}}

In the past several MicroCAT structures have been investigated
\cite{Amir3}. Recently, an optimised MicroCAT mesh has been produced
by Stork Screens\footnote{Stork Screens, Boxmeer, Netherlands}. In
this section we compare the charge transfer derived from
current measurements to simulations, and we
demonstrate that the new device is superior to previously
investigated MicroCAT structures with respect to maximum electron
transparency (to maximise the effective gain) and minimum ion feedback
(to avoid field perturbations in the drift region).

\subsection{Characteristics}
\begin{figure}
 \vspace{0mm}
 \begin{center}
  \includegraphics[clip,width=7.5cm]{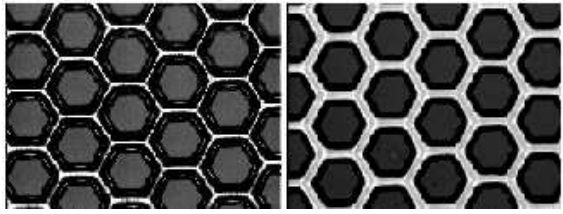}
 \end{center}
  \caption{Microscopic photography of both sides of the optimised
   MicroCAT structure.}
  \label{fig_opmcat}
\end{figure}

\begin{figure}
 \vspace{0mm}
 \begin{center}
  \includegraphics[clip,width=6.5cm]{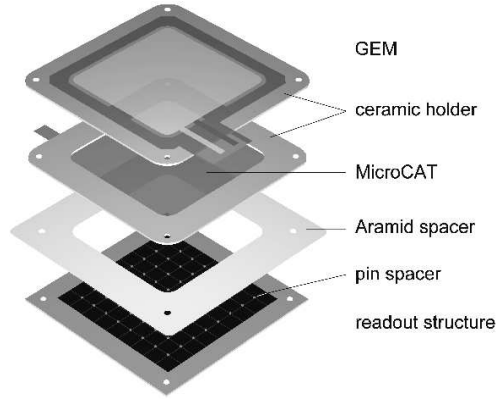}
 \end{center}
  \caption{Schematic view of the detector setup. The distance
 between the GEM and the MicroCAT amounts to
 $2\,\mathrm{mm}$. The drift cathode is mounted at a distance of
 $25\,\mathrm{mm}$ above the GEM structure.}
  \label{fig_gemmcat}
\end{figure}

The MicroCAT structure (see Fig. \ref{fig_opmcat}) has been
optimised with respect to maximum electron transparency and
minimum ion feedback while retaining a good mechanical stability
\cite{Andre1}.
The parameters of the newly built device are given together with those of
previously used MicroCATs in Table \ref{table_MCATtypes}.
\begin{table}
  \begin{center}
  \begin{tabular}{c|c|c|c|c|c}
   type&  holes/inch & pitch $l$ &
         hole diameter $h$ &
         open area & thickness
  \\ \hline \hline
   MCAT155 & 155 & $164 \,\mathrm{\mu m}$ & $116\,\mathrm{\mu m}$ &
   $45.4\,\%$ & $70\, \mathrm{\mu m}$
  \\ 
   MCAT215 & 215 & $118 \,\mathrm{\mu m}$ & $79\,\mathrm{\mu m}$  &
   $40.6\,\%$ & $55\, \mathrm{\mu m}$ 
  \\ 
   MCAT305 & 305 & $83 \,\mathrm{\mu m}$ & $45\,\mathrm{\mu m}$  & 
   $26.5\,\%$ & $55\, \mathrm{\mu m}$ 
  \\
   Opt. MCAT215 & 215 & $118 \,\mathrm{\mu m}$ & $78\,\mathrm{\mu m}$  &
   $39.5\,\%$ & $25\, \mathrm{\mu m}$ 
  \\ 
 \end{tabular}
   \caption{Dimensions of the MicroCAT structures.}
   \label{table_MCATtypes}
   \end{center}
\end{table}
The holes which have a hexagonal shape with edges rounded by the
production process form a hexagonal lattice. 
The cathode distance between the MicroCAT
and the subjacent anode amounts to 
$d=(130\pm10)\,\mathrm{\mu m}$ (see Fig. \ref{fig_gemmcat}),
which is in fact very close to the optimum distance of about
$100\,\mathrm{\mu m}$ with respect to maximum
time-resolution.
The anode is realised by an interpolating
resistive readout structure \cite{Besch,Hendrik1}.
The drift region has a height of $27\,\mathrm{mm}$
(since the GEM structure is not mounted). For all
measurements in this section the
detector volume is filled with a gas mixture of
$1\,\mathrm{bar}$ $\text{Ar/CO}_2$ (90/10). 
\\
The charge transfer behaviour of the MicroCAT depends only on the ratio
$\eta=E_\text{drift}/E_\text{MCAT}$ of the fields above and below the
mesh. This has been confirmed by simulations \cite{Andre1}.

\subsection{Charge transfer behaviour}
For the determination of the charge transfer the measurement of at
least two of the following three currents is neccessary:
\begin{align}
  I_\text{drift} &= I_0 + \delta\cdot\varepsilon
  \cdot(G-1)\cdot I_0 
\label{eq_Idrift}
\\
  I_\text{MCAT} &= -(1-\varepsilon)\cdot I_0 +
  (1-\delta)\cdot\varepsilon\cdot
  (G-1)\cdot I_0 
\label{eq_IMCAT}
\\
  I_\text{anode} &= -\varepsilon\cdot G\cdot I_0
\label{eq_Ianode}
\end{align}
with $I_0$ as the incoming current: 
\begin{align}
  I_0 = R\cdot \frac{E_\gamma}{W}\cdot e\,.
\label{eq_I0}
\end{align}
Here, $\varepsilon$ 
denotes the electron transparency, $\delta$ the
ion feedback, which describes the fraction of ions drifting back to the drift 
cathode, $G$ the gas gain, $R$ the photon rate, $E_\gamma$ the
mean energy deposited by the photons in the detector,
$W$ the mean ionisation potential of the gas atoms/molecules and $e$
the elementary charge.
\\
Since Eqs. \ref{eq_Idrift}-\ref{eq_Ianode} fulfill current
conservation, equivalent to
$\sum I=0$, only two of the three relations are
independent and thus not sufficient to determine all three unknown
variables $\varepsilon$, $\delta$ and $G$.
Therefore, one additional condition is required, which we determine
by the simulation presented in the following paragraph.

\subsubsection{Simulations}

\begin{figure}
  \vspace{0mm}
  \begin{center}
   \includegraphics[clip,width=7.5cm]{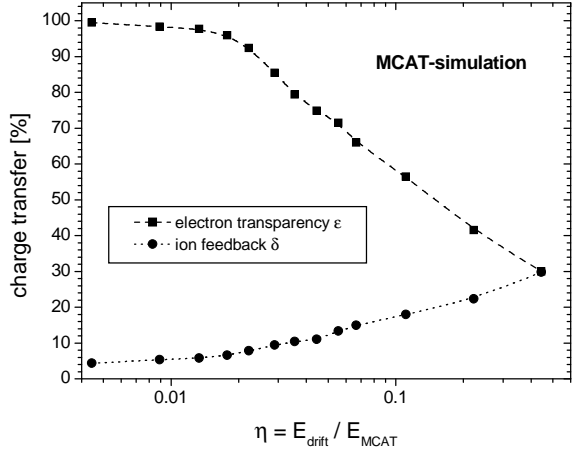}
  \end{center}
  \caption{Simulated electron transparency $\varepsilon$ and ion
  feedback $\delta$ 
  for the optimised MCAT215 structure.}
  \label{fig_mcat_chargetransfersimulation}
\end{figure}

The electric fields for the MicroCAT structures  
have been calculated in three dimensions using the
Maxwell package \cite{Maxwell}. The charge drift was calculated 
using the Garfield program
\cite{Garfield}. The gas properties used for the Garfield simulation
like electron diffusion or drift velocity are calculated by the
Magboltz program \cite{Magboltz2,Magboltz1}.
The distance between the lower side of the MCAT 
and the anode has been set
to $100\,\mathrm{\mu m}$, the MCAT voltage to $-450\,\mathrm{V}$ while
the anode structure has zero potential, leading to an average electric
field in the amplification gap of $E_\text{MCAT}=45\,\mathrm{kV/cm}$.
All potentials are assumed to be constant during the simulation since
the change of the potential during one multiplication process due to
charge motion in the real detector is calculated to be in the order of
$\mathcal{O(\mathrm{mV})}$.
\\
Diffusion has been included for the drifting electrons but neglected for
the ions since this contribution is expected to be small. 
Charge multiplication has not been considered.
The ion feedback simulation is based on an effective size of the ion
cloud of about $\sigma=30\,\mathrm{\mu m}$. A more detailed
description of the charge transfer simulation can be found in
Ref. \cite{Andre1}.
\\
The results of the simulation for the electron transparency
$\varepsilon$ and ion feedback $\delta$ 
are shown in 
Fig. \ref{fig_mcat_chargetransfersimulation}
as a function of the ratio $\eta=E_\text{drift}/E_\text{MCAT}$ for
fixed $E_\text{MCAT}$. 
As expected, the ion feedback increases with
$\eta$ but the electron transparency decreases. 

\subsubsection{Measurements and comparison with simulations
\label{MCATmeasurements}}

\begin{figure}
  \vspace{0mm}
  \begin{center}
   \includegraphics[clip,width=7.5cm] {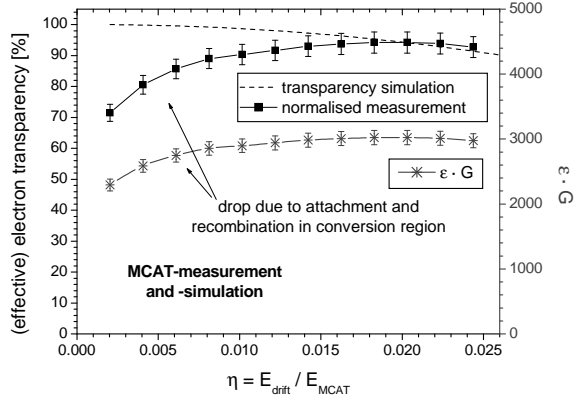}
  \end{center}
  \caption{Measured product $\varepsilon\cdot G$ 
  and comparison of 
  the normalised product with the simulated electron
  transparency for the optimised MCAT215 structure.}
  \label{fig_mcat_epsiloncomparison}
\end{figure}

\begin{figure}
  \begin{center}
   \includegraphics[clip,width=7.5cm] {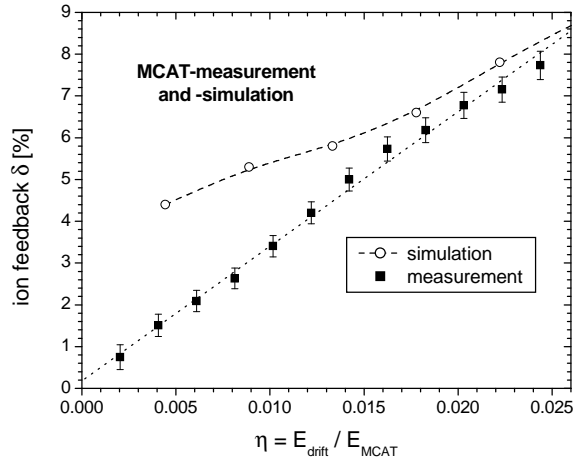}
  \end{center}
  \caption{Comparison of the measured and simulated ion feedback
   $\delta$
   as a function of $\eta$ for the optimised MicroCAT structure.}
  \label{fig_mcat_deltacomparison}
\end{figure}

We have measured the electric currents at all electrodes as a function
of the ratio $\eta=E_\text{drift}/E_\text{MCAT}$ with the settings
$E_\text{MCAT}\approx49\,\mathrm{kV/cm}$ and $d=130\,\mathrm{\mu m}$.
Due to technical
reasons the measurements of $\varepsilon$ and
 $\delta$ are restricted
to small values of $\eta$ below about $0.03$. In Section
\ref{combination} the range of $\eta$ is extended to larger values of
about $0.16$.
\\
From the measured currents we compute the effective gain
$G_\text{eff}$ using the
relations 
\ref{eq_Idrift}--\ref{eq_Ianode}:
\begin{align}
  G_\text{eff}=\varepsilon\cdot G=\frac{I_\text{drift} +
  I_\text{MCAT}}{I_0}\,.
\label{eq_transparency1}
\end{align}
The gain variation due to the increasing drift field in the
investigated range of $\eta$ has been
calculated with Garfield for electrons starting $100\,\mathrm{\mu m}$ 
above the MCAT along the symmetry axis of a hole to be less than
$2\,\%$. Therefore, the measured product $\varepsilon\cdot G$ is expected to
follow the simulation of $\varepsilon$. 
In Fig. \ref{fig_mcat_epsiloncomparison}
where we compare the measurement with the simulation, we observe that
contrary to the simulation the measured product drops at small values of
$\eta$. We attribute this effect to attachment of the primary
electrons and to recombination with back drifting positive ions.
These effects which have not been considered in the simulation seem to be
negligible at higher drift fields corresponding to $\eta>0.02$. We
derive an effective electron transparency $\varepsilon$ (see
Fig. \ref{fig_mcat_epsiloncomparison}) by normalising the measured
$\varepsilon\cdot G$ 
to the simulation of $\varepsilon$ at
$\eta=0.02$. The systematic uncertainty of this procedure is indicated
in the figure by the enlarged error bars.
\\
We also compare the ion feedback $\delta$ with the measurement using the
reasonable approximation $G\gg1$:
\begin{align}
  \delta
  \approx
  \frac{I_\text{drift}-I_0}{I_\text{drift} + I_\text{MCAT}}\,.
\label{eq_ionfb}
\end{align}
The measurement deviates considerably from the simulation at small
$\eta$ illustrated in
Fig. \ref{fig_mcat_deltacomparison}.
Obviously, our model is too coarse in this region of $\eta$.

\subsubsection{Comparison with other MicroCAT types}

\begin{figure}
  \begin{center}
   \includegraphics[clip,width=7.5cm] {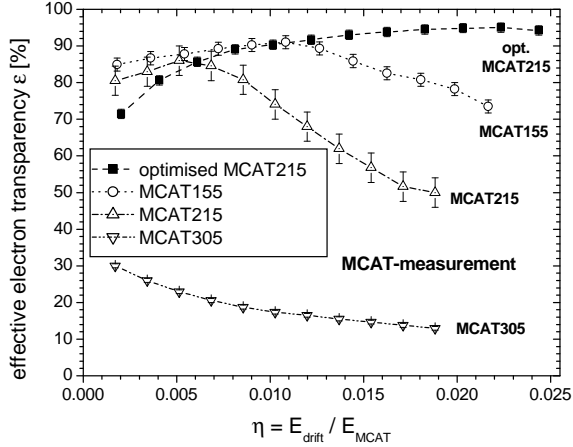}
  \end{center}
  \caption{Comparison of the measured effective electron transparency 
  $\varepsilon$ of four  
  MicroCAT structures as a function of $\eta$.}
  \label{fig_mcat_comparison1}
\end{figure}

\begin{figure}
  \begin{center}
   \includegraphics[clip,width=7.5cm] {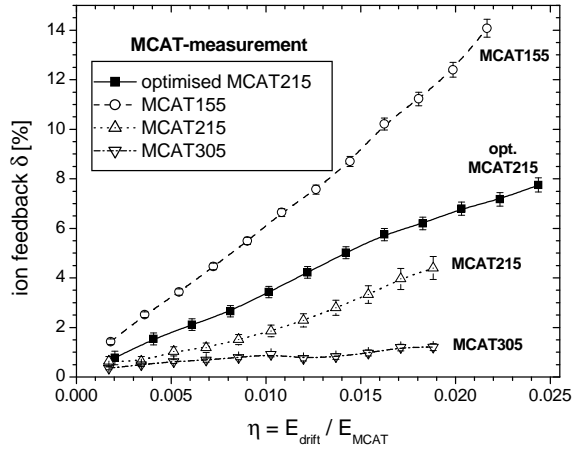}
  \end{center}
  \caption{Comparison of the measured ion feedback
  $\delta$ of four  
  MicroCAT structures as a function of $\eta$.}
  \label{fig_mcat_comparison2}
\end{figure}

Fig. \ref{fig_mcat_comparison1}
shows a comparison of the measured effective 
electron transparency $\varepsilon$ of four MicroCAT
structures with dimensions summarised in Table \ref{table_MCATtypes}.
The optimised mesh offers the largest electron transparency. 
The drop of the effective 
electron transparency at low $\eta$ is different for 
the different MicroCATs because the amount of gas impurities slightly varies
from measurement to measurement. 
\\
The comparison of the ion feedback $\delta$ 
of the MicroCAT structures
(see Fig. \ref{fig_mcat_comparison2})
shows, that the fraction of back drifting ions of the 
optimised MCAT215 structure is
nearly as small as for the old MCAT215 structure and lower than that
of the MCAT155 device, although the electron transparency $\varepsilon$
of the optimised mesh is much larger than that of the competing
devices (compare to Fig. \ref{fig_mcat_comparison1}).

\subsection{Charge multiplication behaviour}

\begin{figure}
  \begin{center}
   \includegraphics[clip,width=7.5cm] {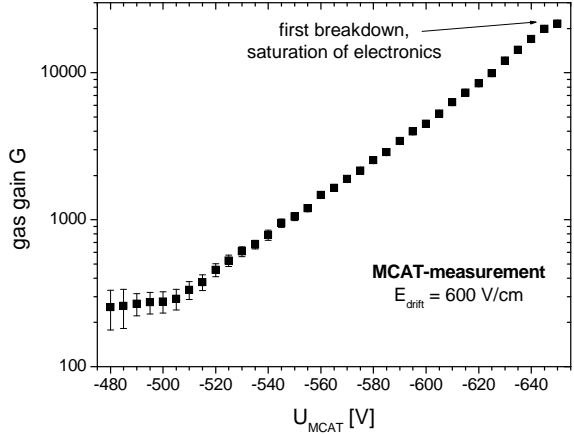}
  \end{center}
  \caption{Measured gas gain $G$ as a function of the applied MCAT voltage.}
  \label{fig_mcat_gasgain}
\end{figure}

Knowing the transparency $\varepsilon$,
the gas gain $G$ can be calculated from Eq. \ref{eq_transparency1}:
\begin{align}
  G=\frac{I_\text{drift} + I_\text{MCAT}}{\varepsilon\cdot I_0}\,.
\label{eq_gasgain}
\end{align}
Fig. \ref{fig_mcat_gasgain}
shows the gain $G$ as a function of the MCAT voltage. It has been
obtained from the observed pulse heights of an $^{55}\text{Fe}$-source
where the gain had been calibrated at one voltage to the current
measurement. The maximum gain, which we define as the gas
amplification, that can be reached within a discharge limit of less
than $1\,\text{spark}/30\,\mathrm{s}$, is in the order of $2\cdot10^4$.

\subsection{Conclusion}

The comparison of the optimised and the three previously investigated
MicroCAT structures shows that the new mesh is superior with respect
to electron transparency. The ion feedback of
the optimised structure is in a reasonable range ($<8\,\%$ up to 
$\eta=0.025$).
In order to reach maximum effective gain the optimised device
should be operated with ratios $\eta\approx0.02$ of the fields above
and below the MCAT.
A maximum gain of about $2\cdot10^4$ can
be obtained at atmospheric pressure for Argon gas mixtures.

\section{Study of the GEM
\label{SingleGEM}}

The charge transfer and gas gain behaviour of the GEM is investigated by
simulations and current measurements to obtain the optimum operation
parameters in view of a combination of the GEM with the MicroCAT. The
influence of different voltage settings and gas mixtures and pressures
has been studied. 

\subsection{Characteristics}

\begin{figure}
  \begin{center}
   \includegraphics[clip,width=4cm] {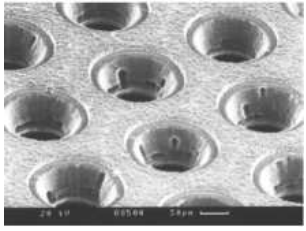}
  \end{center}
  \caption{Scanning electron microscopic photo of the GEM.}
  \label{fig_gemmodel2}
\end{figure}

\begin{figure}
  \begin{center}
   \includegraphics[clip,width=7.5cm] {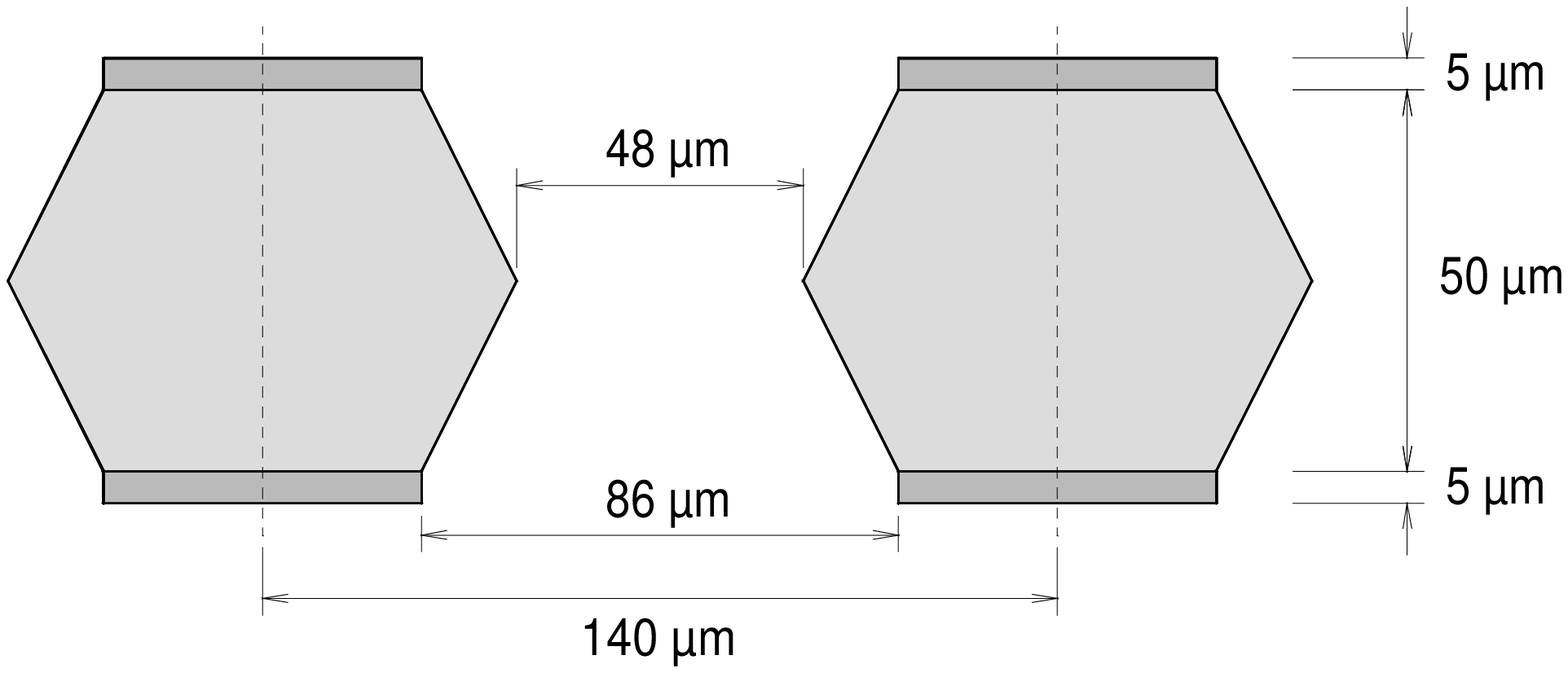}
  \end{center}
  \caption{Vertical cut through the GEM with the dimensions 
   used for the simulation.}
  \label{fig_gemmodel}
\end{figure}

The GEM structure is fixed above the MicroCAT at a
distance of $2\,\mathrm{mm}$ (see Fig. \ref{fig_gemmcat}). 
The MicroCAT serves as anode with a slightly positive voltage 
($U_\text{MCAT}\approx+5\,\mathrm{V}$). In this geometry
the drift region above the
GEM amounts to about $25\,\mathrm{mm}$, the
region below the gas electron multiplier corresponds to the distance
to the MicroCAT structure, consequently $2\,\mathrm{mm}$. 
We denote the electric field above and below the GEM
with drift field and transfer field, respectively.
\\
The GEM \cite{Bennloch} (see Fig. \ref{fig_gemmodel2}), 
itself, has a hexagonal hole arrangement with a
pitch of $140\,\mathrm{\mu m}$, an optical transparency of about
$12\,\%$ and a total
thickness of $60\,\mathrm{\mu m}$ (see Fig. \ref{fig_gemmodel}).
\\
The current measurements have been
carried out with a
$^{55}\text{Fe}$-source, collimated to $5\,\mathrm{mm}^2$,
emitting photons with a rate of about $45\,\mathrm{kHz}$.

\subsection{Charge transfer behaviour
\label{GEMchargetransfer}}

The charge transfer behaviour of the GEM can be described by 
four current equations. Two electron transparencies are introduced:
$\varepsilon_1$ is the probability, that the electrons, coming
from the drift/conversion region, reach the centre of a GEM hole;
$\varepsilon_2$ is
the probability, that these electrons reach the anode. 
This approach has already been introduced by other groups (see for
example in Ref. \cite{Richter}).
The ion feedback is again denoted by $\delta$.
\begin{align}
  I_\text{drift} &= I_0 + \delta\cdot\varepsilon_1\cdot(G-1)\cdot I_0 
\label{eq_GIdrift}
\\
  I_\text{GEM-top} &= -(1-\varepsilon_1)\cdot I_0 + 
  (1-\delta)\cdot\varepsilon_1\cdot(G-1)\cdot I_0 
\label{eq_GIGEMtop}
\\
  I_\text{GEM-bottom} &= -\varepsilon_1\cdot(1-\varepsilon_2)\cdot
  G\cdot I_0 
\label{eq_GIGEMbottom}
\\
  I_\text{anode} &= -\varepsilon_1\cdot\varepsilon_2\cdot G\cdot I_0
\label{eq_GIanode}
\end{align}
Current conservation reduces these equations to three independent relations.
In the following we make use of the effective gain which is defined by
$G_\text{eff}=\varepsilon_1\cdot\varepsilon_2\cdot G$.

\subsubsection{Simulations}

Like in the previous section the three-dimensional 
electric field in the GEM geometry
and the charge transfer behaviour have been investigated with Maxwell 
and Garfield, respectively. For the transparency simulation
the charge multiplication process has
not been considered; ion feedback simulations have not been carried
out.
\\
If not stated differently, we have assumed in the simulations a drift
field of $E_\text{drift}=500\,\mathrm{V/cm}$, a transfer field of
$E_\text{trans}=2000\,\mathrm{V/cm}$, a GEM voltage of
$\Delta U_\text{GEM}=450\,\mathrm{V}$ and a gas mixture of
$1\,\mathrm{bar}$ and $2.5\,\mathrm{bar}$
$\text{Ar/CO}_2$ (90/10).

\begin{figure}
  \begin{center}
   \includegraphics[width=9cm] {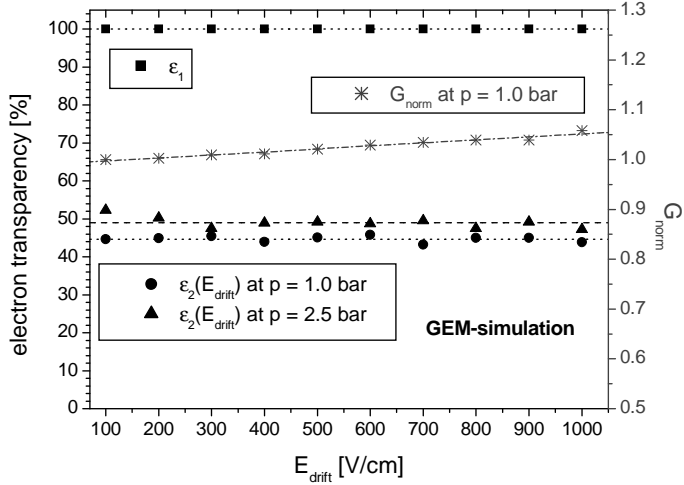}
  \end{center}
  \caption{Simulated electron transparencies $\varepsilon_1$ and
   $\varepsilon_2$  and normalised gain 
  as a function of the drift field.}
  \label{fig_gem_driftsim1}
\end{figure}

\paragraph{Influence of the drift field:}
The dependency of the electron transparencies  $\varepsilon_1$ and
$\varepsilon_2$ on the applied drift field 
($100\,\mathrm{V/cm}\le E_\text{drift}\le1000\,\mathrm{V/cm}$)
is shown in 
Fig. \ref{fig_gem_driftsim1}.
All electrons
reach the centre of the hole in the GEM
structure ($\varepsilon_1=1$), independent of the chosen gas pressure.
The parameter
$\varepsilon_2$ is not influenced noticeably by the drift field.
The deviation of $\varepsilon_2$ between the two gas pressures can be
explained by the smaller electron diffusion at higher pressure.
\\
The situation is very different from that in particle physics
applications where usually due to rather high drift fields
$\varepsilon_1$ is smaller than $100\,\%$. In our case the overall
electron transparency is obviously limited by diffusion effects which
cause a large fraction of electrons to be caught by the lower
electrode.
\\
We have also studied the effect of drift field variations on the gain $G$.
The gain was calculated with Garfield for one electron
starting $100\,\mathrm{\mu m}$ above the GEM along the symmetry axis
of a hole. 
The smallest gain $G_\text{norm}(E_\text{drift})$, which is obtained
at a drift field of $E_\text{drift}=100\,\mathrm{V/cm}$, is normalised to
$1$. The gain 
increases by about $5\,\%$ in the investigated drift field
range. 

\begin{figure}
 \begin{center}
   \includegraphics[width=9cm] {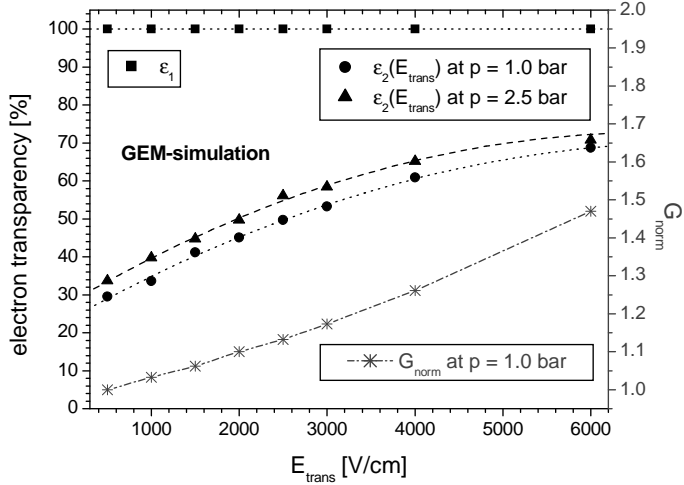}
  \end{center}
  \caption{Simulated electron transparencies  $\varepsilon_1$ and
   $\varepsilon_2$ and normalised gain
  as a function of the transfer field. 
  }
  \label{fig_gem_driftsim2}
\end{figure}

\paragraph{Influence of the transfer field:} The
electron transparencies $\varepsilon_1$ and $\varepsilon_2$ as a
function of the applied transfer field 
($500\,\mathrm{V/cm}\le E_\text{trans}\le6000\,\mathrm{V/cm}$)
are shown in Fig. \ref{fig_gem_driftsim2}.
The transparency $\varepsilon_1$ is constant $\approx100\,\%$ 
for all applied transfer
fields, whereas $\varepsilon_2$ shows a strong increase with rising transfer
fields.
Smaller electron diffusion at higher pressure leads to a larger
$\varepsilon_2$. 
Fig. \ref{fig_gem_driftsim2} also shows the normalised gain
$G_\text{norm}$ as a function  of the transfer field. The change at
low transfer fields is about $5\,\%/(1000\,\mathrm{V/cm})$; 
identical as observed
with a variation of the drift field (see Fig. \ref{fig_gem_driftsim1}).
At transfer fields $E_\text{trans}\gtrsim4000\,\mathrm{V/cm}$ 
the parallel plate
amplification starts in the gap between the GEM and the anode plane, 
resulting in an additional contribution to the gain $G$.

\begin{figure}
  \begin{center}
   \includegraphics[width=9cm] {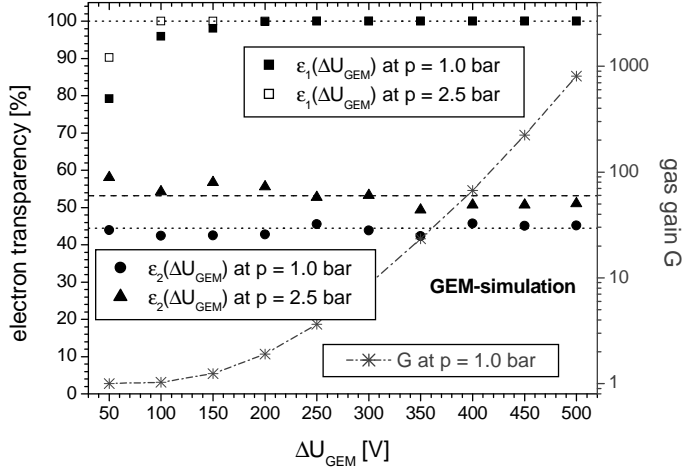}
  \end{center}
  \caption{Simulated electron transparencies $\varepsilon_1$ and
   $\varepsilon_2$ and gas gain as a function of
   the GEM voltage.}
  \label{fig_gem_driftsim3}
\end{figure}

\paragraph{Influence of the GEM voltage:}
The dependence of the electron transparencies 
on the applied GEM voltage
($50\,\mathrm{V}\le \Delta U_\text{GEM}\le500\,\mathrm{V}$)
has been studied (see Fig. \ref{fig_gem_driftsim3}).
The transparencies $\varepsilon_1$ and $\varepsilon_2$ depend only weakly
on the GEM voltage. For very low GEM voltages $\varepsilon_1$
decreases because the drift channel through the GEM holes becomes wider.
Smaller diffusion at a gas pressure of $2.5\,\mathrm{bar}$ results 
in a larger electron transparency than for atmospheric pressure. 
The calculated GEM gain $G$ has the typical exponential
shape with increasing operation voltage.

\subsubsection{Measurements and comparison with simulations}

A gas mixture of $\text{Ar/CO}_2$ (90/10) at a pressure of
$1\,\mathrm{bar}$ was chosen for most of the measurements
if not stated differently. In addition, the charge
transfer and multiplication behaviour was investigated for several
Argon, Krypton, Xenon and $\text{CO}_2$-quench gas mixtures at various
pressures up to $2.5\,\mathrm{bar}$. 
\\ 
From Eqs. \ref{eq_GIdrift}--\ref{eq_GIanode}
the product $\varepsilon_1\cdot G$, the transparency
$\varepsilon_2$ and the ion feedback $\delta$ (using the reasonable 
approximation $G\gg1$) can be calculated: 
\begin{align}
  \label{eq_gemgain1}
  \varepsilon_1\cdot G&=-\frac{(I_\text{GEM-bottom}+I_\text{anode})}{I_0}
\end{align}
\begin{align}
  \label{eq_gemgain2}
  \varepsilon_2 &=\frac{I_\text{anode}}{I_\text{GEM-bottom}+I_\text{anode}}
\end{align}
\begin{align}
  \label{eq_gemgain3}
  \delta&\approx\frac{I_\text{drift}-I_0}{I_\text{drift}+I_\text{GEM-top}}
\end{align}

\begin{figure}
  \begin{center}
   \includegraphics[clip,width=7.5cm] {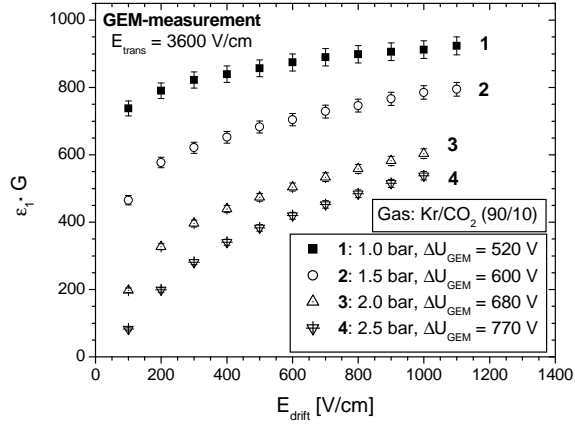}
  \end{center}
  \caption{Measured 
  product $\varepsilon_1\cdot G$ as a function of the drift
  field for several gas pressures of $\text{Kr/CO}_2$ (90/10).}
  \label{fig_gem_eps1g5}
\end{figure}

\begin{figure}
  \noindent
 \unitlength=1mm
 \makebox[\textwidth][c]{
  \begin{picture}(80,65)
   \put(0,0){\hspace{0mm}\includegraphics[clip,width=7.5cm]
   {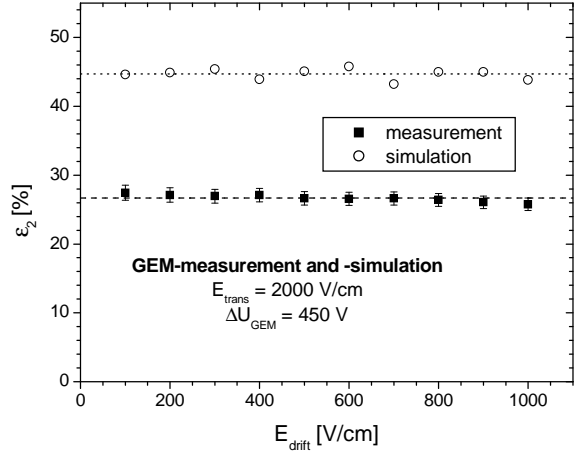}}
  \end{picture}}
  \caption{Comparison between the measured and the simulated
  electron transparency $\varepsilon_2$ as a function of the drift
  field.}
  \label{fig_gem_eps27}
\end{figure}

\begin{figure}
  \noindent
 \unitlength=1mm
 \makebox[\textwidth][c]{
   \begin{picture}(80,65)
   \put(0,0){\hspace{0mm}\includegraphics[clip,width=7.5cm]
   {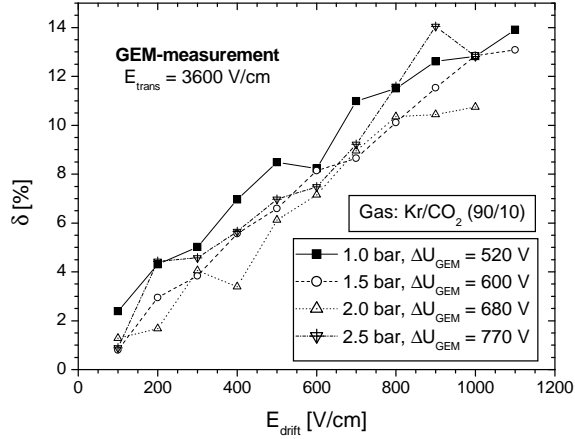}}
  \end{picture}}
  \caption{Measured ion feedback $\delta$ as a function of the drift field
  at different gas pressures of $\text{Kr/CO}_2$ (90/10).}
  \label{fig_gem_delta4}
\end{figure}

\paragraph{Influence of the drift field:\label{driftfield}}
Fig. \ref{fig_gem_eps1g5}
shows the influence of the drift field on the product
$\varepsilon_1\cdot G$ for various gas pressures of $\text{Kr/CO}_2$
(90/10). The decrease at small drift fields is again attributed to
recombination and attachment effects which are expected to increase
with gas pressure. The simulation (see Fig. \ref{fig_gem_driftsim1})
indicates that the gain $G$ slightly depends on the drift field, whereas the
transparency $\varepsilon_1=1$ is constant. Therefore the measured
electron transparency $\varepsilon_1$ can be assumed to
be very close to $100\,\%$.
\\
The measurements show that the transparency $\varepsilon_2$ does not
change noticeably with the varying drift field 
(see Fig. \ref{fig_gem_eps27}).
This is in line with expectation and simulation. However, the
transparency $\varepsilon_2$ is apparently overestimated in the
simulation, which means that in reality more electrons move to the
GEM-bottom electrode. Possible explanations of this effect could be:
\begin{enumerate}
\item{neglection of the avalanche development in the simulation: 
repulsive forces and UV-photons widen the electron cloud.}
\item{underestimation of the electron diffusion at high electric
fields in Magboltz.}
\end{enumerate}
The measured ion feedback $\delta$ of the GEM is presented in 
Fig. \ref{fig_gem_delta4}
for several gas pressures of $\text{Kr/CO}_2$ (90/10).
It increases nearly linearly with rising drift field.

\begin{figure}
  \begin{center}
   \includegraphics[clip,width=7.5cm] {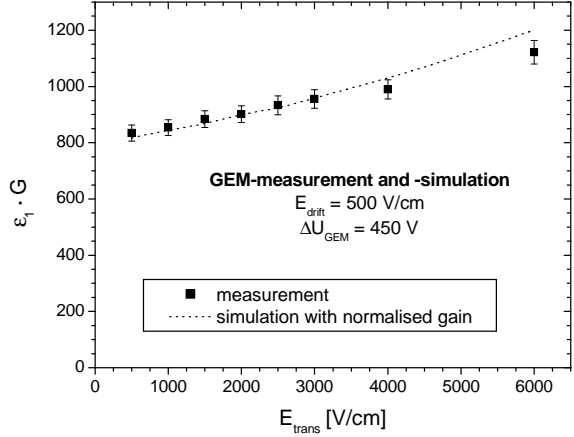}
  \end{center}
  \caption{Comparison between the measured and the simulated
  product $\varepsilon_1\cdot G$ as a function of the transfer
  field.}
  \label{fig_gem_eps1g8}
\end{figure}

\begin{figure}
  \noindent
 \unitlength=1mm
 \makebox[\textwidth][c]{
   \begin{picture}(80,65)
   \put(0,0){\hspace{0mm}\includegraphics[clip,width=7.5cm]
   {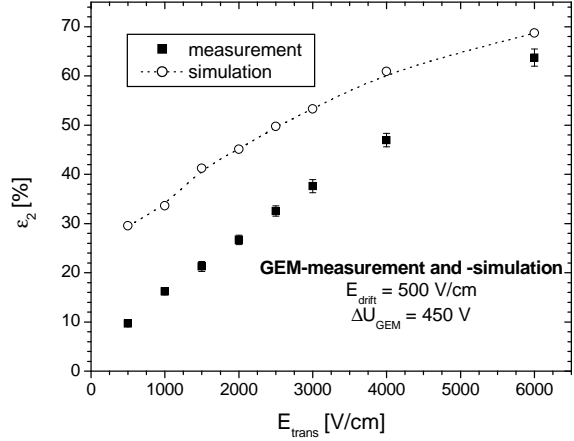}}
  \end{picture}}
  \caption{Comparison between the measured and the simulated
 transparency $\varepsilon_2$ as a function of the transfer
  field.}
  \label{fig_gem_eps28}
\end{figure}

\begin{figure}
  \noindent
 \unitlength=1mm
 \makebox[\textwidth][c]{
   \begin{picture}(80,65)
   \put(0,0){\hspace{0mm}\includegraphics[clip,width=7.5cm]
   {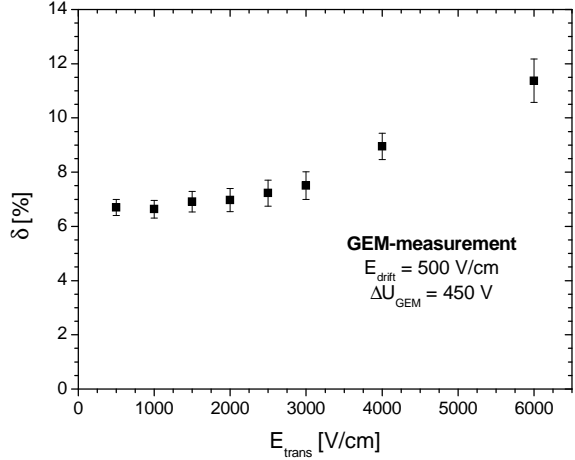}}
  \end{picture}}
  \caption{Measured ion feedback $\delta$ as a function of the transfer
  field.}
  \label{fig_gem_delta2}
\end{figure}

\paragraph{Influence of the transfer field:}
The measured increase of $\varepsilon_1\cdot G$ with rising
$E_\text{trans}$ (see Fig. \ref{fig_gem_eps1g8})
is in a good agreement with the increase of the gain $G$ which was
predicted by the simulation (compare to
Fig. \ref{fig_gem_driftsim2}). 
Thereby the simulated 
gain had to be renormalised because of the inaccurate knowledge of the
corresponding Townsend coefficients. 
We conclude that
$\varepsilon_1$ is not sizeably influenced by the transfer field.
\\
However, the transparency $\varepsilon_2$ increases strongly with 
rising transfer fields (see Fig. \ref{fig_gem_eps28}).
The deviation of the absolute values of $\varepsilon_2$ between simulation and
measurement has already been discussed in section \ref{driftfield}.
For transfer fields
$E_\text{trans}\gtrsim4000\,\mathrm{V/cm}$ the parallel plate amplification
underneath the GEM
leads to an apparent rise of $\varepsilon_2$ which is more pronounced
in the measurement due to the longer multiplication path of
$2\,\mathrm{mm}$ compared to $100\,\mathrm{\mu m}$ in the simulation.
\\
The measured ion feedback $\delta$ (see Fig. \ref{fig_gem_delta2})
is approximately constant at lower transfer fields. When the parallel
plate amplification produces a noticeable amount of gain (at large
transfer fields) the ion feedback rises. We conclude that the ion
feedback is larger for ions which are produced below the GEM rather
than in the GEM holes.

\begin{figure}
  \noindent
 \unitlength=1mm
 \makebox[\textwidth][c]{
   \begin{picture}(80,65)
   \put(0,0){\hspace{0mm}\includegraphics[clip,width=7.5cm]
   {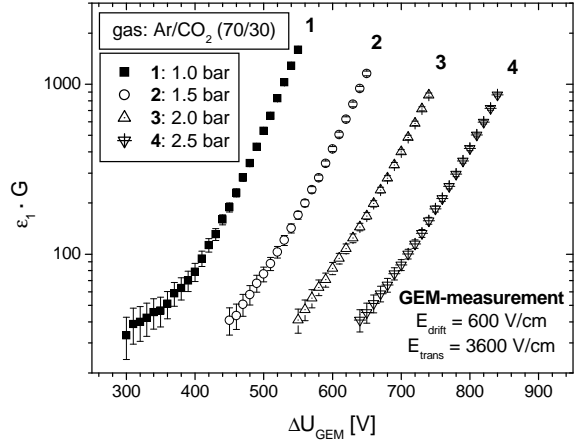}}
  \end{picture}}
  \caption{Measured product 
  $\varepsilon_1\cdot G$ as a function of the GEM
  voltage in $\text{Ar/CO}_2$ (70/30) at 
  several gas pressures.}
  \label{fig_gem_eps1g6}
\end{figure}

\begin{figure}
  \noindent
 \unitlength=1mm
 \makebox[\textwidth][c]{
   \begin{picture}(80,65)
   \put(0,0){\hspace{0mm}\includegraphics[clip,width=7.5cm]
   {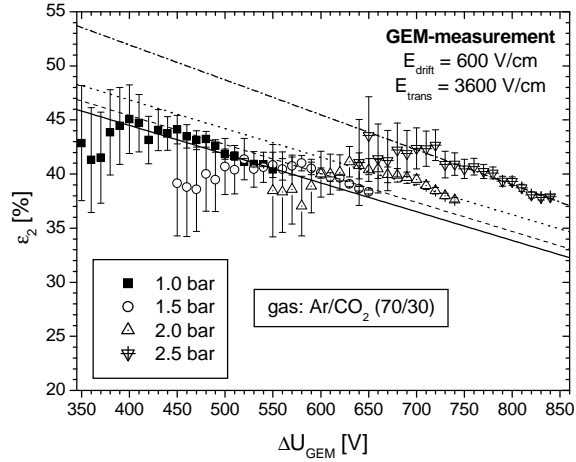}}
  \end{picture}}
  \caption{Measured electron transparency
  $\varepsilon_2$ as a function of the GEM
  voltage in $\text{Ar/CO}_2\,(70/30)$ at 
  several gas pressures.}
  \label{fig_gem_eps26}
\end{figure}

\begin{figure}
  \noindent
 \unitlength=1mm
 \makebox[\textwidth][c]{
   \begin{picture}(80,65)
   \put(0,0){\hspace{0mm}\includegraphics[clip,width=7.5cm]
   {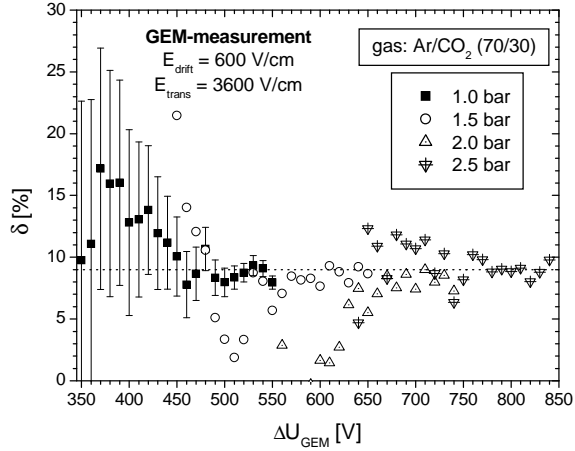}}
  \end{picture}}
  \caption{Measured ion feedback $\delta$ as a function of the GEM
  voltage in $\text{Ar/CO}_2\,(70/30)$ at 
  several gas pressures.}
  \label{fig_gem_delta5}
\end{figure}

\paragraph{Influence of the GEM voltage:}
The expected exponential dependence of
$\varepsilon_1\cdot G$ on the GEM voltage is shown in
Fig. \ref{fig_gem_eps1g6}
for $\text{Ar/CO}_2$ (70/30) at various gas pressures.
\\
With increasing $\Delta U_\text{GEM}$ the transparency $\varepsilon_2$
decreases (see Fig. \ref{fig_gem_eps26}).
For high GEM fields the electron drift channel becomes very narrow. As
a consequence the probability that electrons diffuse to one of the
dense drift lines that end on the GEM bottom electrode is increased.
\\
The ion feedback as a function of the GEM voltage is shown
in Fig. \ref{fig_gem_delta5}.
Due to very small currents at the drift cathode at low gains
the error bars are very large. The deviations from a constant
behaviour are not significant.

\subsection{Charge multiplication behaviour}
\begin{figure}
  \noindent
 \unitlength=1mm
 \makebox[\textwidth][c]{
  \begin{picture}(80,65)
   \put(0,0){\hspace{0mm}\includegraphics[clip,width=7.5cm]
   {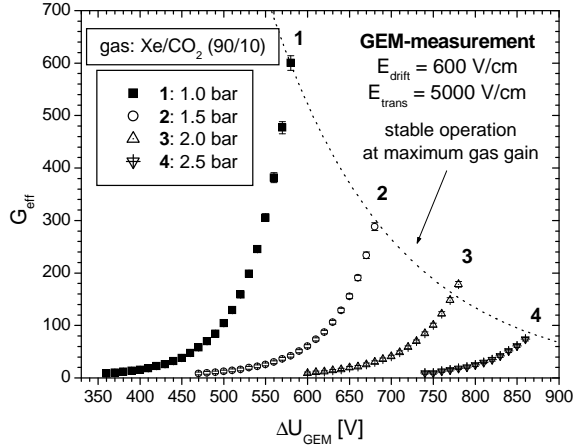}}
  \end{picture}}
  \caption{Measured effective gain
  $G_\text{eff}=\varepsilon_1\varepsilon_2 G$ as a function of the
  GEM voltage in $\text{Xe/CO}_2$ (90/10) at serveral gas pressures.}
  \label{fig_gem_geff1}
\end{figure}

We have investigated the gain behaviour of the GEM for
different gas mixtures and pressures.
The maximum gas amplification,
which can be reached within a discharge limit of less than 
$1\,\text{spark}/30\,\mathrm{s}$, decreases drastically with gas
pressure 
(shown as an example for a $\text{Xe/CO}_2$ (90/10) gas mixture  
in Fig. \ref{fig_gem_geff1}).
This effect is studied in more detail in Ref. \cite{Bondar}.

\subsection{Conclusion}

The approach of the introduction of the two electron transparencies
$\varepsilon_1$ and $\varepsilon_2$ has been sensibly confirmed by our
measurements.
\\
The systematic studies of the currents at the electrodes in a wide 
range of voltages and electric fields show the following results:

\begin{itemize}
\item{The transparencies $\varepsilon_1$ and $\varepsilon_2$ do not
depend significantly on the drift field for
$E_\text{drift}\lesssim1\,\mathrm{kV/cm}$. The ion feedback $\delta$
rises nearly linearly with increasing drift field. In order to avoid
electron losses due to 
recombination and attachment in the conversion region the drift field
should not be chosen too small, especially for higher gas
pressures. The gain $G$ rises slightly with increasing drift field.}
\item{The transfer field has no influence on the
electron transparency $\varepsilon_1$ in the investigated field range
$E_\text{trans}\lesssim6\,\mathrm{kV/cm}$. The
transparency $\varepsilon_2$ strongly rises with increasing transfer
fields. The ion feedback $\delta$ does not change noticeably for small
transfer fields and increases slightly for higher transfer fields.
Also the gain $G$ rises with increasing transfer field.}
\item{Very small GEM voltages lead to a decrease of
$\varepsilon_1$. The transparency $\varepsilon_2$ decreases slightly
for large potential differences, whereas the ion feedback $\delta$
is nearly constant.}
\end{itemize}

\section{The combination of the optimised MicroCAT with the GEM
\label{combination}}

The combination of the optimised MCAT215 
and the GEM has been investigated by current
measurements to determine the optimum voltage settings with respect to
maximum effective gain and minimum ion feedback. The maximum
gas gain in different gas environments has beed studied.
No simulations have been carried out. However, the results of the
individual simulations from Sections \ref{MCAT} and \ref{SingleGEM}
can be combined.

\subsection{Characteristics}
Fig. \ref{fig_gemmcat} shows the schematic setup of the combination of 
MicroCAT and GEM. We call the electric field between the drift 
cathode and the GEM by $E_\text{drift}$, the
field between the GEM and the MCAT by
$E_\text{trans}$ and between the MCAT and the anode by
$E_\text{MCAT}\equiv U_\text{MCAT}/d$, where $U_\text{MCAT}$ denotes
the voltage applied to the MCAT and $d=130\,\mathrm{\mu m}$ 
the distance between MCAT and anode. 
The ratio of the fields above and below the MicroCAT
structure is denoted by $\eta=E_\text{trans}/E_\text{MCAT}$. 
The electron transparency, ion feedback and gain of the MicroCAT 
are denoted by $\varepsilon_\text{MCAT}$, $\delta_\text{MCAT}$ and
$G_\text{MCAT}$, respectively. 
All GEM transparencies are denoted as in the previous section, the
GEM gain is denoted by $G_\text{GEM}$. The
total effective gain is defined as follows:
\begin{align}
  G_\text{eff}=\varepsilon_1 \cdot\varepsilon_2\cdot
\varepsilon_\text{MCAT}\cdot G_\text{GEM}\cdot G_\text{MCAT}
\label{eq_MGGeff}
\end{align}

\subsection{Charge transfer behaviour}

\begin{figure}
  \noindent
 \unitlength=1mm
 \makebox[\textwidth][c]{
   \begin{picture}(80,65)
   \put(0,0){\hspace{0mm}\includegraphics[clip,width=7.5cm]
   {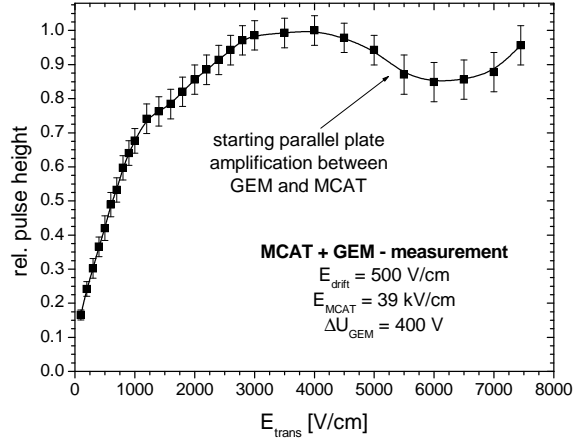}}
  \end{picture}}
  \caption{Measured relative pulse height as a function of the
 transfer field $E_{\text{trans}}$ for the combination of MCAT
 and GEM.}
  \label{fig_mcatgem_relpulseheight1}
\end{figure}

\begin{figure}
  \noindent
 \unitlength=1mm
 \makebox[\textwidth][c]{
   \begin{picture}(80,65)
   \put(0,0){\hspace{0mm}\includegraphics[clip,width=7.5cm]
   {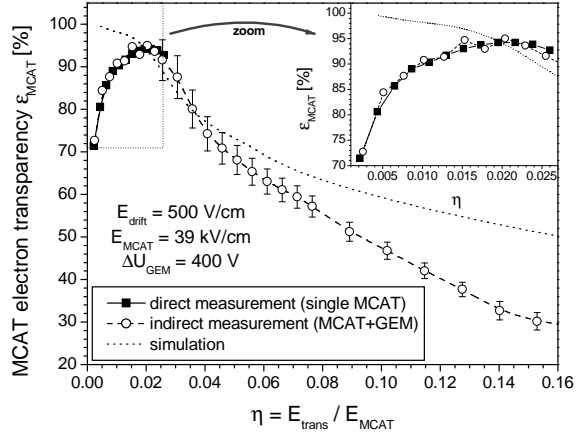}}
  \end{picture}}
  \caption{Comparison between the directly and indirectly determined
 electron transparency $\varepsilon_\text{MCAT}$ as a function of
 $\eta$ for constant $E_\text{MCAT}$. 
 In addition the simulated electron transparency is shown.}
  \label{fig_mcatgem_relpulseheight3}
\end{figure}

\begin{figure}
  \noindent
 \unitlength=1mm
 \makebox[\textwidth][c]{
   \begin{picture}(80,65)
   \put(0,0){\hspace{0mm}\includegraphics[clip,width=7.5cm]
   {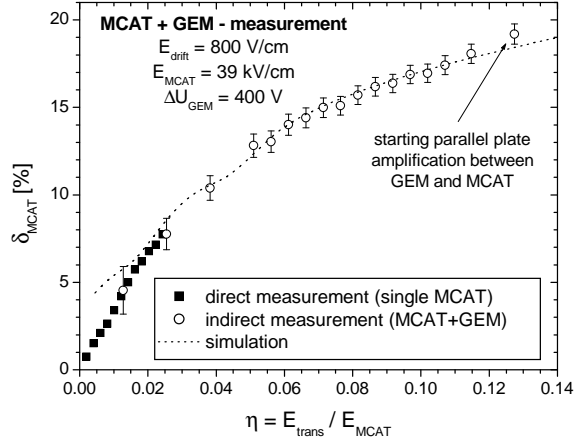}}
  \end{picture}}
  \caption{Comparison between the directly and indirectly determined
 ion feedback $\delta_\text{MCAT}$
 as a function of $\eta$ for constant $E_\text{MCAT}$.
 In addition the simulated ion feedback is shown.}
  \label{fig_mcatgem_deltamcat}
\end{figure}

\begin{figure}
  \noindent
 \unitlength=1mm
 \makebox[\textwidth][c]{
   \begin{picture}(80,65)
   \put(0,0){\hspace{0mm}\includegraphics[clip,width=7.5cm]
   {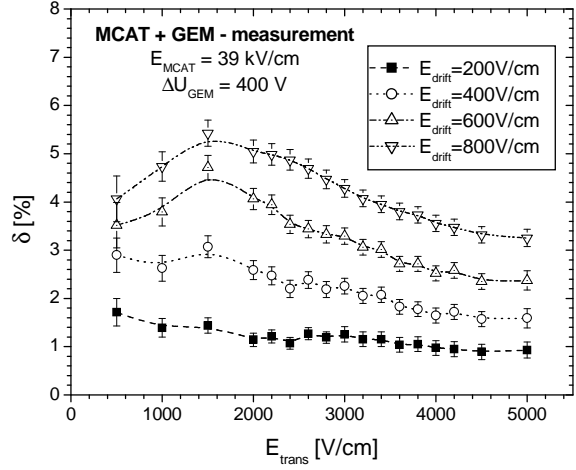}}
  \end{picture}}
  \caption{Measured total ion feedback $\delta$ as a function of the transfer
 field $E_{\text{trans}}$.} 
  \label{fig_mcatgem_deltatotal}
\end{figure}

\begin{figure}
  \noindent
 \unitlength=1mm
 \makebox[\textwidth][c]{
   \begin{picture}(80,65)
   \put(0,0){\hspace{0mm}\includegraphics[clip,width=7.5cm]
   {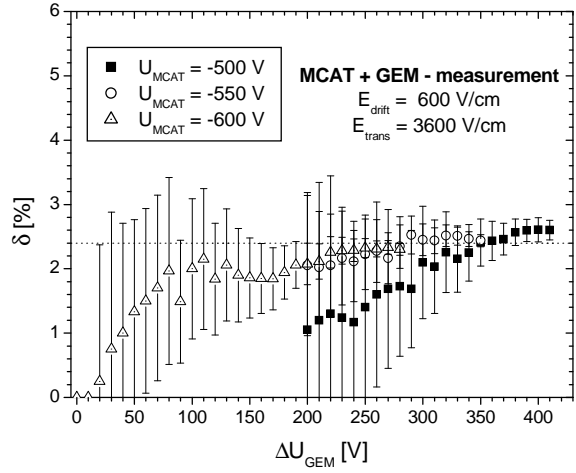}}
  \end{picture}}
  \caption{Measured total ion feedback $\delta$ as a function of the 
  GEM voltage $\Delta U_\text{GEM}$.}
  \label{fig_mcatgem_ionfeedback1}
\end{figure}

\begin{figure}
  \noindent
 \unitlength=1mm
 \makebox[\textwidth][c]{
   \begin{picture}(80,65)
   \put(0,0){\hspace{0mm}\includegraphics[clip,width=7.5cm]
   {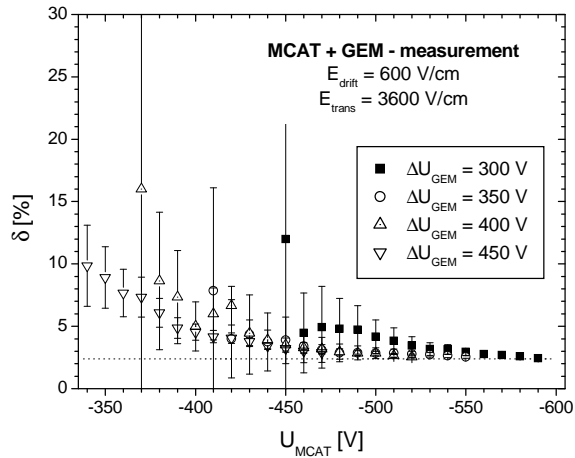}}
  \end{picture}}
  \caption{Measured total ion feedback $\delta$ as a function of the 
  MCAT voltage $U_\text{MCAT}$.}
  \label{fig_mcatgem_ionfeedback2}
\end{figure}

Fig. \ref{fig_mcatgem_relpulseheight1}
shows the relative signal pulse heights as
a function of the transfer field for a drift 
field of $E_\text{drift}=500\,\mathrm{V/cm}$, a GEM
voltage of $\Delta U_\text{GEM}=400\,\mathrm{V}$, an MCAT
voltage of $U_\text{MCAT}=-510\,\mathrm{V}$ at a gas pressure of
$1\,\mathrm{bar}$ $\text{Ar/CO}_2$ (90/10).
Up to transfer fields of $E_\text{trans}\approx3000\,\mathrm{V/cm}$
the signal pulse height rises. The subsequent plateau reaches up to 
$E_\text{trans}\approx4000\,\mathrm{V/cm}$, before the signals
get smaller again. At very high transfer fields of
$E_\text{trans}\gtrsim5000\,\mathrm{V/cm}$ parallel plate
amplification between GEM and MicroCAT 
starts, which again leads to an increase of the signals'
pulse heights.
\\
The shape of this curve is affected by two opposite
effects: on the one hand the transparency 
$\varepsilon_2$ of the GEM rises for increasing transfer fields 
(compare to Fig. \ref{fig_gem_eps28}), and on the other hand
the electron transparency $\varepsilon_\text{MCAT}$ of 
the MicroCAT drops for larger values of $\eta$ (compare to
Fig. \ref{fig_mcat_chargetransfersimulation}). With the knowledge of
$\varepsilon_2(E_\text{trans})$, determined in section \ref{SingleGEM},
the electron transparency of the MCAT can be
calculated up to values of $\eta\approx0.2$.
The result is shown in Fig. \ref{fig_mcatgem_relpulseheight3}.
It agrees perfectly with the transparency measured directly with the
MicroCAT alone (see Fig. \ref{fig_mcat_epsiloncomparison}).
Measurements at different GEM voltages 
show the same results. However, a large deviation from the simulation
is obvious for $\eta\gtrsim0.04$. The measured transparency
$\varepsilon_\text{MCAT}$ is significantly 
smaller than what the simulation predicts. We have no explanation for
this effect.
\\
By means of current relations and the reasonable assumptions, that
$\varepsilon_1=1$, $G_\text{GEM}\gg1$ and $G_\text{MCAT}\gg1$
the ion feedback $\delta_\text{MCAT}$ of the MicroCAT can be determined:
\begin{align*}
  \delta_\text{MCAT}\approx 1-\frac{I_\text{MCAT}}{G_\text{eff}\cdot I_0}-
                        (1-\varepsilon_\text{MCAT})\cdot\varepsilon_2\cdot
                        \frac{G_\text{GEM}}{G_\text{eff}}
\end{align*}
The result (see Fig. \ref{fig_mcatgem_deltamcat})
shows a good agreement to the direct measurement (compare to
Fig. \ref{fig_mcat_deltacomparison}). Despite the
coarse model for the ion feedback simulation the
simulation and the measurement match very well for larger ratios $\eta$.
Additional measurements have shown that
the field in the conversion region
$E_\text{drift}\lesssim1\,\mathrm{kV/cm}$ above the GEM 
has no influence on the ion feedback $\delta_\text{MCAT}$. 
\\
Fig. \ref{fig_mcatgem_deltatotal}
shows, that the total ion feedback 
$\delta = -I_\text{drift}/I_\text{anode}$ is getting more 
favourable for higher transfer fields. The GEM voltage has no strong effect
on the total ion feedback $\delta$ 
(see Fig. \ref{fig_mcatgem_ionfeedback1}).
The increase of the total ion feedback $\delta$ with decreasing
MCAT voltage (see Fig. \ref{fig_mcatgem_ionfeedback2})
can be explained by the rising contribution of the
ion feedback $\delta_\text{MCAT}$, which gets larger
for higher ratios
$\eta=E_\text{trans}/E_\text{MCAT}$ (compare to Fig.
\ref{fig_mcatgem_deltamcat}).

\subsection{Charge multiplication behaviour}

\begin{figure}
  \noindent
 \unitlength=1mm
 \makebox[\textwidth][c]{
   \begin{picture}(80,65)
   \put(0,0){\hspace{0mm}\includegraphics[clip,width=7.5cm]
   {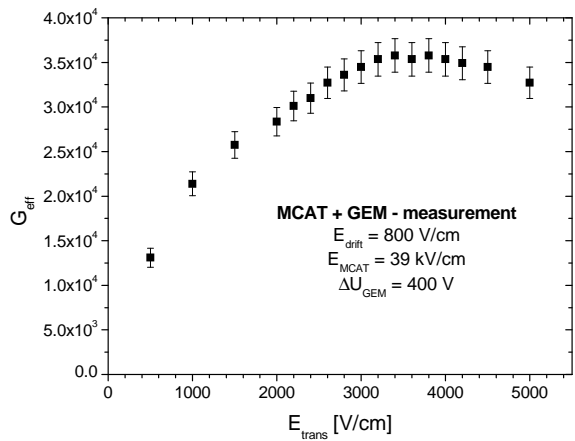}}
  \end{picture}}
  \caption{Measured 
  effective gain $G_\text{eff}$ as a function of the transfer field
  $E_{\text{trans}}$.}
  \label{fig_mcatgem_geff1}
\end{figure}

\begin{figure}
  \noindent
 \unitlength=1mm
 \makebox[\textwidth][c]{
   \begin{picture}(80,65)
   \put(0,0){\hspace{0mm}\includegraphics[clip,width=7.5cm]
   {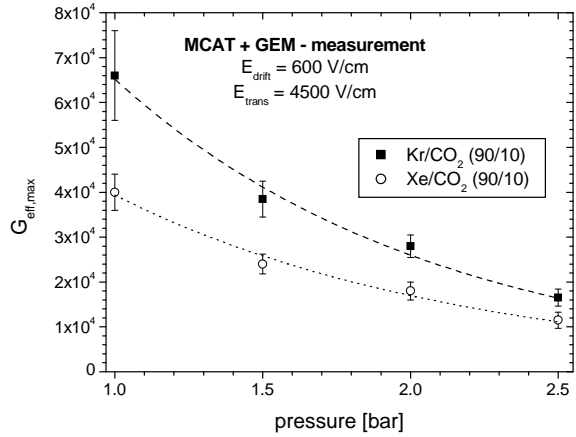}}
  \end{picture}}
  \caption{Maximum gas gain $G_\text{eff,max}$ as a function of the 
  pressure for a $\text{Kr/CO}_2$ (90/10) and a 
  $\text{Xe/CO}_2$ (90/10) gas mixture. The lines
  correspond to exponential fits of the measured data.}
  \label{fig_mcatgem_geffmax1}
\end{figure}

\begin{figure}
  \noindent
 \unitlength=1mm
 \makebox[\textwidth][c]{
   \begin{picture}(80,65)
   \put(0,0){\hspace{0mm}\includegraphics[clip,width=7.5cm]
   {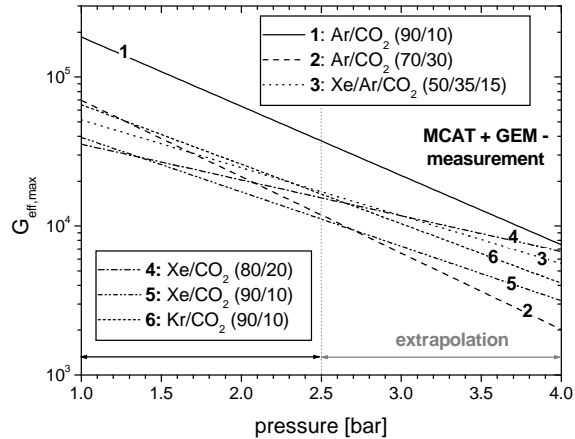}}
  \end{picture}}
  \caption{Maximum gas gain $G_\text{eff,max}$ as a function of the 
  pressure for six different gas mixtures. The curves correspond to
  exponential fits of the measured data and are extrapolated up to
  pressures of $4\,\mathrm{bar}$.}
  \label{fig_mcatgem_geffmax2}
\end{figure}

The effective gain $G_\text{eff}$  (see Eq. \ref{eq_MGGeff}) 
of the GEM/MicroCAT-combination
can be calculated from the measured currents $I_0$ (see Eq. \ref{eq_I0})
and $I_{\text{anode}}$
as follows:
\begin{align*}
  G_\text{eff} &= -\frac{I_\text{anode}}{I_0}
\end{align*}
The effective gain 
as a function of the transfer field rises to
maximum at $E_\text{trans}\approx3.6\,\mathrm{kV/cm}$
in $1\,\mathrm{bar}$ $\text{Ar/CO}_2$ (90/10) 
(see Fig. \ref{fig_mcatgem_geff1}).
\\
In order to study the maximum gas gain
a systematic investigation of the effective gain $G_\text{eff}$ as a
function of the gas pressure has been carried out for 
different noble gas/quench gas
mixtures.
The GEM voltages are chosen such, that this fragile pre-amplification
device always works in a safe range, whereas the MCAT produces the
main amplification. This method has the advantage that all discharges
appear below the MicroCAT structure, which is
not easily damaged by sparks.
All sparks appear at a defined avalanche size depending on the
gas mixture and pressure used, but independent on the voltage settings on
GEM or on MCAT.
\\
The measurements of the maximum gas gain are carried out for
gas pressures up to $2.5\,\mathrm{bar}$. Fig. 
\ref{fig_mcatgem_geffmax1}
shows the measured maximum gain for several pressures of
$\text{Kr/CO}_2$ (90/10) and $\text{Xe/CO}_2$ (90/10). 
In a first approximation the drop of the gain
for higher pressures is approximately exponential.
The results of the exponentially fitted data are shown in 
Fig. \ref{fig_mcatgem_geffmax2}. 
The largest gas gain of about $2\cdot10^{5}$
can be achieved with the gas mixture of
$\text{Ar/CO}_2$ (90/10). At atmospheric
pressure the gain decreases with the atomic number of the nobel gas.
We estimate that for gas pressures up to $4\,\mathrm{bar}$ gains
of at least 
$G_\text{eff,max}>2\cdot10^3$ can be reached, even for Xe-mixtures.

\subsection{Conclusion}

The investigation of the optimum voltage settings for a 
GEM/MicroCAT-combination leads to the
following results:
\begin{itemize}
\item{To avoid recombination and attachment effects 
the drift field above the GEM should be chosen larger than 
$500\,\mathrm{V/cm}$. Higher gas pressure requires larger
drift fields (compare to Section \ref{SingleGEM} and 
Fig. \ref{fig_gem_eps1g5}).}
\item{A transfer field in the range of $3000-4000\,\mathrm{V/cm}$
ensures a maximum effective gain and a small total ion feedback.}
\item{Since the combined gain of the GEM and the MicroCAT is limited
and because the GEM is easily destroyed by discharges the MicroCAT
should be operated at much higher gain than the GEM.}
\end{itemize}
The maximum gas gain decreases with pressure and atomic number of
the noble gas. A stable operation with gains in the order of
$10^4$ can be obtained even for Xenon gas mixtures up to
$2.5\,\mathrm{bar}$. 

\section{Conclusion}

The new optimised MicroCAT structure 
is superior to all other MicroCAT devices with respect to
electron transparency. The ion feedback
of the new mesh is smaller than for the MCAT155 and
nearly as good as for the old MCAT215 structure.
Gas gains of larger than $2\cdot10^4$
can be achieved in Argon gas mixtures at standard pressure.
\\
In our special detector setup a GEM has
been investigated with respect to its charge transfer and
gas gain behaviour with different
voltage settings and different gas environments. The charge
transfer simulations
are well confirmed by the measurements. 
A sub-division of the overall electron transparency into the two
transparencies $\varepsilon_1$ and $\varepsilon_2$ is reasonable.
The achieved gains of the
GEM are comparable to published results.
\\
The combination of both gas gain devices
leads to a very high gas gain at atmospheric pressure and a reliable
detector operation with gains in the order of $10^4$ in Xenon
gas mixtures at higher pressures of about $2.5\,\mathrm{bar}$. The GEM
can be operated in a very safe gas gain mode, whereas the
the MicroCAT produces the largest amount of gain;
due to the robustness of the nickel mesh and its insensitivity against
sparking no destruction of this device needs to be feared. The fraction
of back drifting ions is in the order of a few percent when
MicroCAT and GEM are combined. Therefore, spatial reconstruction 
distortions due to space charge effects in the conversion region are
expected to be small.

\begin{ack}
We are indebted to G. Claassen and T. van de
Mortel from Stork Screens. Only due to their effort
the prototype production of the optimised MicroCAT mesh was possible.
We are grateful to G. Zech for valuable comments that have contributed
to this publication.
\end{ack}

\end{document}